\documentclass[10pt,english,american,letterpaper]{article}
\usepackage[T1]{fontenc}
\usepackage[latin1]{inputenc}
\usepackage{graphicx}
\usepackage{amssymb}

\makeatletter

\makeatletter

\makeatletter

\usepackage{osameet2}

\makeatother

\makeatother

\usepackage{babel}
\makeatother
\begin{document}

\title{Coherence and Correlations in Atom Lasers}

\author{P. D. Drummond$^{1}$, T. Vaughan$^{1}$, J. F. Corney$^{1}$, G.
Leuchs$^{2}$, P. Deuar$^{3,4}$}

\maketitle
\address{
 {$^{1}$ARC Centre of Excellence for Quantum-Atom Optics, The University of Queensland, Brisbane QLD 4072, Australia},\\
 {$^{2}$Max-Planck Forschungsgruppe, Universitat Erlangen-Nurnberg},\\
 {$^{3}$Universit\'{e} Paris-Sud, LPTMS, UMR8626, B\^{a}t. 100, Universit\'{e} Paris-Sud 91405 Orsay cedex, France},\\
 {$^{4}$CNRS, LPTMS, UMR8626, B\^{a}t. 100, Universit\'{e} Paris-Sud 91405 Orsay cedex, France} }

\email{drummond@physics.uq.edu.au}

\begin{abstract}
We review and characterize the quantum coherence measures that are
most useful for quantum gases, including Bose-Einstein condensates
(BEC) and ultra-cold fermions, and outline how to calculate these
in the typically dynamical environment of an interacting multi-mode
quantum gas. 
\end{abstract}

\section{Introduction}

Understanding the coherence and correlations of ultra-cold quantum
gases can be thought of as an extension of Glauber's seminal work\cite{Glauber1,Glauber2,Glauber3}
on the quantum coherence properties of photons. Indeed, the basic
measured quantities are the quantum correlation functions of the gas,\cite{GlauberAO}
just as in the photonic systems treated by Glauber.

We start with the fundamental principle that particle detection experiments
measure n-th order correlations\emph{:}

\begin{equation}
G^{(n)}(\mathbf{x}_{1}\ldots\mathbf{x}_{n},\mathbf{x}_{n+1},\ldots\mathbf{x}_{2n})=\left\langle \hat{\Psi}^{\dagger}\left(\mathbf{x}_{1}\right)\hat{\Psi}^{\dagger}\left(\mathbf{x}_{n}\right)\hat{\Psi}\left(\mathbf{x}_{n+1}\right)\ldots\hat{\Psi}\left(\mathbf{x}_{2n}\right)\right\rangle \end{equation}
 If $\mathbf{x}_{n}=\vec{x}_{n},t_{n}$ , then these normally-ordered
correlations correspond to an absorptive measurement of $n$ particles,
at $n$ distinct space-time events. Nonabsorptive measurements are
simply density correlations\cite{Lukin}, but we will focus on the
normally ordered correlations for definiteness. They then can be used
to define normalized correlation measures:

\begin{equation}
g^{(n)}(\mathbf{x}_{1}\ldots\mathbf{x}_{n},\mathbf{x}_{n+1},\ldots\mathbf{x}_{2n})=\frac{G^{(n)}(\mathbf{x}_{1}\ldots\mathbf{x}_{n},\mathbf{x}_{n+1},\ldots\mathbf{x}_{2n})}{\sqrt{\prod_{i}G^{(1)}(\mathbf{x}_{i},\mathbf{x}_{i})}}\end{equation}

In particular, we wish to ask: what correlations \emph{characterize}
an $N$-particle ultra-cold gas? Given that, in principle, all correlation
functions are measurable, we must focus on those that convey useful
information. We will investigate center-of-mass measurements, condensation
measurements, and pair-correlation properties.

Thus, the main questions of interest here are:

\begin{itemize}
\item Center-of-mass position and momentum

\begin{itemize}
\item What is the `standard quantum limit' for position measurement? 
\item We find that fermions with a given density distribution have a lower
variance than bosons. 
\end{itemize}
\item Measures of condensation into a single mode, or orbital, as in a BEC

\begin{itemize}
\item How can one measure Bose condensation into an unknown single-particle
state? 
\item We show that one should use higher-order correlations to characterize
a condensate in general, under conditions that the condensed mode
is unknown and may have statistical fluctuations! 
\end{itemize}
\item Dynamical correlations in quantum nonlinear \textbf{atom} optics

\begin{itemize}
\item How can one calculate statistical correlations that vary in time,
in quantum dynamical experiments? 
\item We show that non-classical phase-space representations provide powerful
methods for these calculations. 
\end{itemize}
\end{itemize}

\section{Ultra-cold Atom Experiments }

The simplicity of ultracold quantum gases is a vital factor in modern
developments in this field. Theoretical descriptions are able to use
simple models\cite{Dalfovo,Leggett}, combining coherence and many-body
theory. The underlying interactions are well characterized by a few
parameters, interactions can be tuned, and new (possibly macroscopic)
tests of quantum mechanics are possible\cite{Macro}.

Recent developments include the well known experiments that have led
to atom lasers, atomic diffraction, and interferometers. However,
a quiet revolution in measurement techniques has also taken place,
leading to the direct detection of atom correlations. This allows
us to ask what quantifies the location, coherence and correlations
of an ultra-cold gas at the quantum noise level, with an assurance
that the corresponding measurements are not impossible.

Thus, the driving force behind theoretical efforts to characterize
the coherence and correlation properties of ultra-cold quantum gases
is the rapid growth in experimental techniques, for both generating
ultra-cold gases and carrying out the requisite correlation measurements.

\begin{figure}
\noindent \begin{centering}\includegraphics[width=7cm]{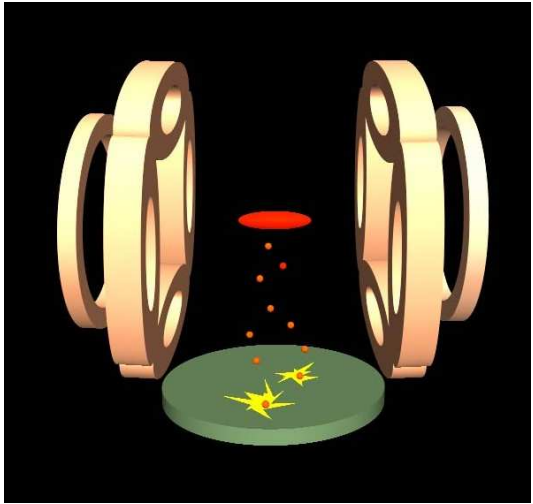} \par\end{centering}

\caption{\label{fig:He* diagram} Schematic diagram of metastable Helium experiments.}
\end{figure}

The first technique for measuring atomic correlations in cold gases
was the pioneering Hanbury-Brown-Twiss two-atom correlation experiment
carried out by Yasuda and Shimizu\cite{Shimizu} on an ultra-cold
metastable $^{20}{\rm Ne}^{*}$ atomic beam. More recently, the development
of multi-channel plate ${\rm He}^{*}$ atom counting techniques\cite{Schellekens2005}
now allows the direct measurement of space and time-resolved arrivals
of either fermonic or bosonic metastable Helium atoms on a detector
plate, as shown in Fig (\ref{fig:He* diagram}). While these methods
still have some limitations with regard to efficiency and time-resolution,
they are in principle able to count almost every atom in a metastable
ultra-cold gas.

The first technique for measuring atomic correlations \emph{inside}
a non-metastable Bose condensate was the use of ultra-cold collisions
as a loss-mechanism, proportional to the local correlation function.
This can be used either directly, or in combination with photo-association
methods. It typically provides spatially integrated information, averaged
over a trap volume. As an example, this method was used to measure
$g^{(3)}$using three-body collisions\cite{Tolra-NIST-exp}, thus
providing the first evidence for the correlations predicted\cite{KK-DG-PD-GS-2003}
in the Tonk-Girardeau regime in a one-dimensional Bose gas.

Another technique of great interest is the use of optical probes.
Such measurements are sensitive to the local density fluctuations,
and were historically used first to study liquids and particles in
suspension via light scattering. Modern techniques typically focus
on forward scattering, and have been used to measure nonlocal density
correlations between fermion pairs\cite{Jinfermi,Fermiantibunch}
produced in molecular `downconversion' experiments, analogous to parametric
downconversion in optics, and also between bosonic atoms\cite{Greiner2005,Folling2005,EsslingerBEC,Elongated-quasi-condensate}.
Other, more exotic correlation measurement techniques are also under
development\cite{Haunanotube,Raizen}.

\begin{figure}
\noindent \begin{centering}\includegraphics[width=7cm]{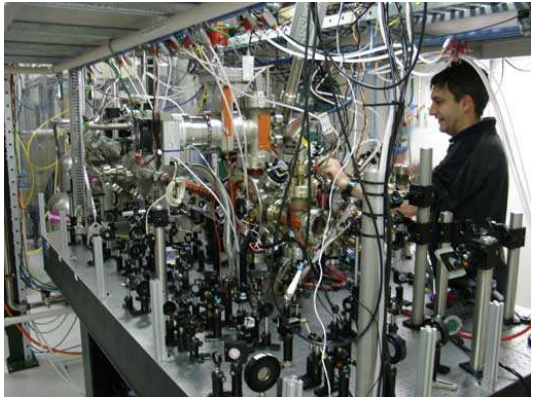} \par\end{centering}

\caption{\label{fig:He* photo} Tabletop setup for metastable Helium experiment.}
\end{figure}

While undoubtedly more complex than a simple BEC-formation experiment,
all of these different atom correlation experiments are still table-top
compatible, as shown in Fig (\ref{fig:He* photo}), which illustrates
a compact metastable Helium correlation experiment at the Australian
National University.

\section{Center of mass measures}

In ultra-cold atom experiments, the BEC or degenerate Fermi gas is
not homogeneous, because of the trap, and is therefore spatially localized.
This introduces the first idea we wish to explore: what is the standard
quantum limit\cite{Vaughan} for a center-of-mass (COM) position and
momentum measurement? The relevant Hermitian operators, for the case
of a number state with eigenvalue $N>0$, are:\begin{eqnarray}
\hat{\mathbf{x}}^{(N)} & = & \frac{\int\mathbf{x}\hat{n}(\mathbf{x})d^{D}\mathbf{x}}{N}\,,\label{eq:COM}\\
\hat{\mathbf{P}} & = & \hbar\int\mathbf{k}\hat{n}(\mathbf{k})d^{D}\mathbf{k},\nonumber \end{eqnarray}
 where $\hat{n}(\mathbf{x})=\hat{\Psi}^{\dagger}\left(\mathbf{x}\right)\hat{\Psi}\left(\mathbf{x}\right)$.
Here the spatial dimension $D$ is included to allow treatment of
reduced dimensional environments. These collective operators obey
the usual commutation relations:

\begin{equation}
[\hat{x}_{j}^{(N)},\hat{P}_{k}]=i\hbar\delta_{jk}\label{eq:comm}\end{equation}
 and Heisenberg Uncertainty Principle:\begin{equation}
\Delta x_{i}^{(N)}\Delta P_{i}\geq\frac{\hbar}{2}\label{eq:HUP}\end{equation}

When there is an uncertainty in the total particle number, a useful
approximation (for large numbers of particles) is to replace the eigenvalue
by the mean operator expectation value:

\begin{equation}
\hat{{X}}=\frac{1}{\langle\hat{N}\rangle}\int\mathbf{x}\hat{n}(\mathbf{x})d^{D}\mathbf{x}\label{eq:COMapprox}\end{equation}

\subsection{Standard Quantum Noise Levels }

We wish to find the characteristic or `standard' quantum noise level
(SQNL) for the COM position and momentum variance, in an idealized
low-noise state. This is analogous to the vacuum noise level of a
quadrature measurement. That is, while it is not the lowest noise
level possible - which is zero - it is a reasonable goal for an experiment
in which technical and thermal noise are eliminated as far as possible.
We consider a SQNL defined relative to given density distribution
$\langle\hat{n}(\mathbf{x})\rangle$, as the COM variance of the ground-state
system configuration in a non-interacting gas with the same density.

Given this definition as the characteristic noise level, we wish to
compare the cases of Fermi and Bose gases. Now the SQNL as defined
here does not require the gas to be trapped - the gas could be in
free space. However, the reference system we choose for comparison
purposes is a trapped, non-interacting gas for simplicity. We note
that it is possible to have the system in a minimum uncertainty state,
which satisfies Eq (\ref{eq:HUP}) replaced by an equality. Quantum
states that are analogous to squeezed states are also possible, in
which there is a trade-off between reduction in one variance, and
increase in the other.

Experimentally, the existence of excess noise from the initial atomic
source, a moving trap potential, residual thermal noise or even from
the quantum noise of the cooling process itself, can result in excess
momentum injected into the atomic cloud. This will cause micro-motion
in the trap, and hence a COM variance for both the position and momentum
which is above the standard limit. Repeated measurements on an ensemble
will result in different values of position and momentum, for each
measurement, with the possible outcome that both types of variance
are well above the SQNL.

However, the simple question we will investigate here is the value
of the variance in the case of a non-interacting ground-state configuration.

\subsection{SQNL - Identical Bosons}

We introduce the notation of \begin{equation}
\hat{a}_{j}=\int\hat{\Psi}\left(\mathbf{x}\right)u_{j}^{*}\left(\mathbf{x}\right)d^{D}\mathbf{x}\,,\end{equation}
 where $u_{j}\left(\mathbf{x}\right)$is lowest single-particle energy
eigenstate of the nominal potential used to obtain the required density
distribution. 

\begin{figure}
\noindent \begin{centering}\includegraphics[width=10cm]{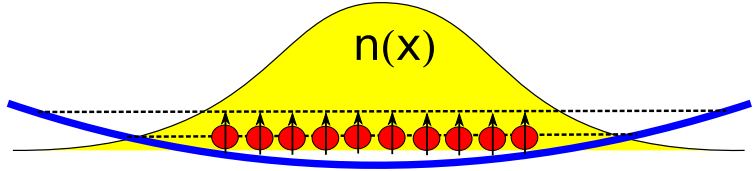} \par\end{centering}

\caption{\label{fig:SQNLBosonsdiagram} Identical bosons in their ground state.}
\end{figure}

The reference ground state, corresponding to $N$ identical bosons
in the single-particle ground state (shown schematically in Fig \ref{fig:SQNLBosonsdiagram})
is then:

\begin{equation}
|\Psi^{(N)}\rangle=\frac{1}{\sqrt{n!}}\left[\hat{a}_{0}^{\dagger}\right]^{N}|0\rangle,\end{equation}
 for which the SQNL for mean position and momentum in any dimension
is: \begin{eqnarray}
\sigma_{Sx}^{2} & = & \langle\Delta\hat{x}_{i}^{2}\rangle=\sigma_{x}^{2}/N\nonumber \\
\sigma_{Sp}^{2} & = & \langle\Delta\hat{P_{i}}^{2}\rangle=N\sigma_{p}^{2}\end{eqnarray}
 Here $\sigma_{p}$ is the standard deviation of the density distribution
in momentum space, equal to $\hbar/(2\sigma_{x})$ if the distribution
is Gaussian (and thus in a minimum uncertainty state).

\subsection{SQNL - Identical Fermions}

\begin{figure}
\noindent \begin{centering}\includegraphics[width=10cm]{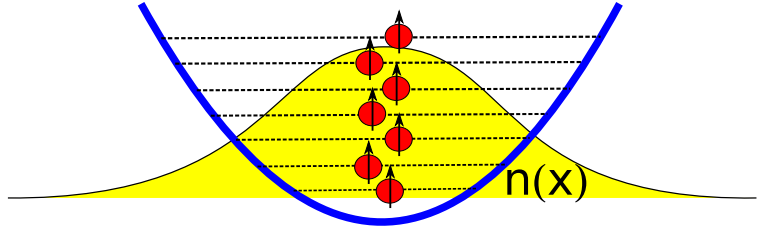} \par\end{centering}

\caption{\label{fig:SQNLFermionsdiagram} Identical harmonically trapped fermions
in their ground state.}
\end{figure}

We can express the zero-temperature state of a collection of $N$
identical fermions using a similar notation:

\begin{equation}
|\Psi^{(N)}\rangle=\left[\prod_{j=1}^{N}\hat{a}_{j}^{\dagger}\right]|0\rangle\end{equation}
 where $j$ labels the $N$ lowest energy eigenstates of some trapping
potential, as shown in figure \ref{fig:SQNLFermionsdiagram}. For
simplicity here, we assume the potential $V\left(\mathbf{x}\right)=\hbar\omega^{2}\left|\mathbf{x}\right|^{2}/2$
is harmonic, with a trap-frequency $\omega$.

The dimensionality of the calculation in this case influences the
result through the mode structure of the ground state of $N$ identical
fermions. For a Fermi gas confined to $D$ spatial dimensions, the
SQNL in the mean position and total momentum is

\begin{eqnarray}
\sigma_{Sx}^{2} & = & \langle\Delta\hat{x}_{i}^{2}\rangle=\sigma_{x}^{2}/N^{1+1/D}\nonumber \\
\sigma_{Sp}^{2} & = & \langle\Delta\hat{P_{i}}^{2}\rangle=N^{1-1/D}\sigma_{p}^{2}\:.\end{eqnarray}
 We thus find the interesting result that the SQNL in measurements
of the mean position is a factor of $\sqrt[D]{N}$ smaller for harmonically
trapped fermions in comparison with the same number of identical bosons
- ie., $N$ times smaller in one dimension - provided the density
variance itself is comparable. We attribute this difference to the
Pauli exclusion principle and the associated Fermi pressure which
expands the density distribution (for the same trap frequency), but
with correlated density fluctuations so that the mean COM variance
is unchanged. In the Fermi case, the expanded size means that $\sigma_{p}\sim N^{1/D}\hbar/(2\sigma_{x})$,
so that the density distribution variances are no longer at the minimum
uncertainty level. However, the minimum uncertainty limit for COM
position and momentum is (approximately) achieved for a harmonic trap
in any number of dimensions.

Surprisingly, it is the characteristic length of the trap ground-state
(divided by $\sqrt{N}$) that sets the absolute COM position variance
scale for either type of quantum statistic.

\section{Condensation measures}

The archetypal criterion for Bose-Einstein condensation in a gas of
interacting identical bosons is the one devoloped in 1956 by Penrose
and Onsager \cite{PenroseOnsager}. Their criterion hinges on the
fact that eigenvalues of the single particle reduced density matrix
$\rho^{(1)}(\mathbf{x},\mathbf{x}')=\langle\hat{\psi}^{\dagger}(\mathbf{x}')\hat{\psi}(\mathbf{x})\rangle$
are the mean occupancies of the corresponding eigenstates. Penrose
and Onsager therefore suggested that the existance of BEC is synonymous
with at least one of these eigenvalues growing extensively with the
size of the system. This is equivalent to requiring perfect first
order (phase) coherence, a fact which is reflected in Glauber's extension
of his theory of optical coherence \cite{Glauber1,Glauber2,Glauber3}
to the ultra-cold gas arena \cite{GlauberAO}.

The question we now ask is this: is first order coherence the defining
feature of a BEC? Recent literature has begun to acknowlege that the
answer may in fact be no. For example, Wilkin et al.~\cite{Wilkin}
have found the zero-temperature ground state of an harmonically trapped
attractively interacting rotating Bose gas is a fragmented condensate
according to the first order criterion. Pethick and Pitaevskii \cite{PethickPitaevskii}
have leveled criticism at this diagnosis, suggesting that the Penrose-Onsager
criterion gives sensible results only when interpreted in the centre-of-mass
frame of the system. This same result has also been considered in
depth by Gajda \cite{Gajda} who suggests that Wilkin et al.'s unintuitive
result might be avoided if the reduced density operator in the Penrose-Onsager
criterion were replaced with `conditional' reduced density operators
-- again effectively tracing out any mean position uncertainty present
in the system.

Additionally, the first-principles evaporative cooling simulations
discussed later in Sec.~\ref{sec:Simulations-of-quantum} provide
further evidence of the short-comings of using first-order coherence
as the basis for a criterion. They indicate that at least one method
of BEC preparation implicitly results in significant mean position
uncertainty which washes out the uniform phase coherence.

\subsection{Probabilistic mixtures of condensed modes}

Examples of states which are handled incorrectly by the Penrose-Onsager
criterion can be easily constructed. For instance, consider the following
mixed state:

\begin{equation}
\hat{\rho}_{M}=\sum_{i}p_{i}|(u_{i})^{N}\rangle\!\langle(u_{i})^{N}|\end{equation}
 where each $|(n_{i})^{N}\rangle=(\hat{a}_{i}^{\dagger N}/\sqrt{N!})|0\rangle$
describes a condensed $N$-particle system with the order parameter
$u_{i}(\mathbf{r}).$ This is therefore a mixture of condensates in
different modes: in every case there is a BEC present.

Such mixtures occur when the center-of-mass motion is excited in a
trapped BEC. This motion is undamped, and observed in almost all experiments.
It may, under some circumstances, result in a relatively low average
occupation of any single mode. This of course leads to the decay of
first-order coherence and thus the condensate fraction as defined
by Penrose and Onsager.

We posit, however, that statistical uncertainty as to the exact mode
of condensation should not influence the classification of that system
as condensed, given the particles deterministically occupy a single
mode.

\subsection{Higher-order condensation measures}

One obvious solution to this problem is to consider measures based
on correlations of order greater than unity. For a set of orthogonal
modes $u(\mathbf{r})$ we define the $m$-th order phase-space filling
factor

\begin{equation}
F_{u}^{(m)}=\frac{\sum_{u}\langle:\hat{n}_{u}^{m}:\rangle}{\langle:\hat{N}^{m}:\rangle}.\end{equation}
 This is a well-defined, obversable correlation function for the Bose
gas that describes the (inverse) spread across the modes $u(\mathbf{r})$
in individual members of a statistical ensemble. It has the property
that it is sensitive to off-diagonal number correlations such as $\langle:\hat{n_{u}}\hat{n}_{u'}\hat{n}_{n''}\ldots:\rangle$
in such a way that the filling factor approaches unity for systems
in mixed states composed only of macroscopically occupied modes in
this particular basis.

The filling factor is therefore a signature of condensation even when
the BEC is oscillating or moving randomly, as it usually is in an
experiment, provided the possible condensate modes belong (or are
`close' to) the set of orthogonal modes present in the numerator of
$F^{(m)}$. .For example, consider the second-order phase-space filling
factor in momentum space:\begin{equation}
F_{\mathbf{k}}^{(2)}=\frac{\sum_{\mathbf{k}}\langle:\hat{n}_{\mathbf{k}}^{2}:\rangle}{\langle:\hat{N}^{2}:\rangle}\label{eq:F2}\end{equation}
 which attains its maximum of one for a state with all of the particles
in a single momentum mode, or for a mixture of such states with possibly
different values of momentum. This is calculable for models of evaporative
cooling and gives a high occupation even for unknown modes.

In terms of spatial correlations, this particular measure can be written
as\foreignlanguage{american}{\begin{eqnarray}
F^{(2)} & = & \frac{1}{V\langle:\hat{N}^{2}:\rangle}\int d\mathbf{x_{1}}d\mathbf{x_{2}}d\mathbf{r}\langle\hat{\Psi}^{\dagger}(\mathbf{x_{1}}-\mathbf{r})\hat{\Psi}^{\dagger}(\mathbf{x_{2}}+\mathbf{r})\hat{\Psi}(\mathbf{x_{2}})\hat{\Psi}(\mathbf{x_{1}})\rangle\nonumber \\
 & = & \frac{1}{V\langle:\hat{N}^{2}:\rangle}\int d\mathbf{x_{1}}d\mathbf{x_{2}}d\mathbf{r}G^{(2)}(\mathbf{x}_{1}-\mathbf{r},\mathbf{x}_{2}+\mathbf{r},\mathbf{x}_{2},\mathbf{x}_{1}),\end{eqnarray}
 }which is just the spatial average of a second-order nonlocal correlation
function. Thus the phase-space filling factor, when defined in terms
of the momentum modes, is a measure of the overall spatial coherence
of the state. But unlike first-order measures of coherence, it is
robust in the presence of mixtures.

\section{Simulations of quantum dynamics}

\label{sec:Simulations-of-quantum}While defining relevant correlation
functions is both instructive and relatively simple, the same cannot
be said for their calculation. We are faced with the problem of how
to calculate an interacting many-body quantum state. Furthermore,
for ultracold atoms, since there is no coupling to a finite-temperature
reservoir, and external potentials and fields are readily adjustable,
most experiments are typically non-equilibrium in nature.

The underlying difficulty is that quantum many-body problems are exponentially
complex. If we consider $n$ atoms distributed among $m$ modes, and
take $n\simeq m\simeq500,000$, then the corresponding number of distinct
quantum states is:

\begin{equation}
N_{S}\approx2^{2n}\approx2^{1,000,000}\end{equation}
 This means that there are more quantum states than atoms in the universe,
so we certainly can't numerically diagonalize the Hamiltonian. Even
if we could find the eigenstates, there are so many that this is not
of itself particularly useful. Calculating $N_{S}$ expansion coefficients
for an arbitrary initial state is necessary, in order to \textbf{use}
the eigenstates in a dynamical calculation. This is not a practical
endeavour.

As an illustration, Fig (\ref{fig:EvapCooldiagram}) gives a diagram
of the collision physics that leads to evaporative cooling. However,
a typical experiment has not just two, but rather millions of initial
atoms (or more). To discover the final quantum density matrix requires
one to integrate forward in time a many-body master equation that
includes both quantum collisions and trap losses. This can be carried
out using a Boltzmann or quantum kinetic equation approximation\cite{GardZoller},
but this is usually at some cost in accuracy. While all stochastic
approaches sacrifice information, what matters is that no systematic
errors are introduced.

\begin{center}%
\begin{figure}[h]
\noindent \begin{centering}\includegraphics[width=8cm]{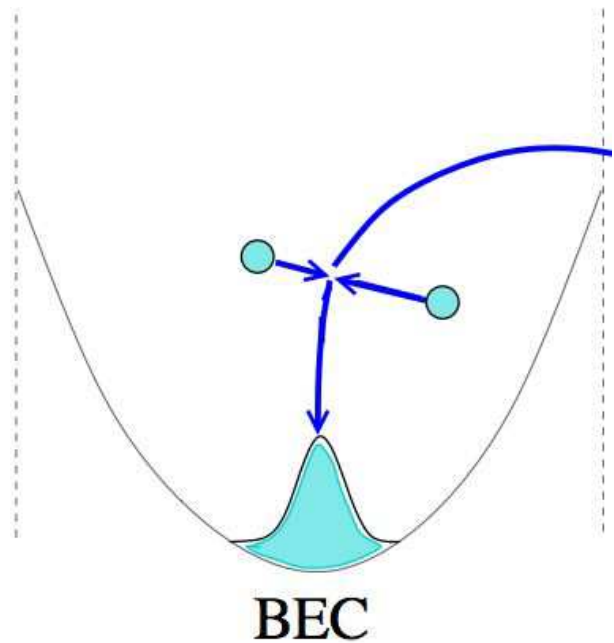} \par\end{centering}

\caption{\label{fig:EvapCooldiagram}Schematic diagram of evaporative cooling
collision}
\end{figure}
\par\end{center}

\subsection{Quantum phase-space methods}

There are a number of phase-space representations of quantum mechanics.
However, if a classical phase-space is used, as in the Wigner\cite{Wigner}
and Glauber-Sudarshan\cite{GSp} representations, with $2M$ degrees
of freedom $(x_{i},p_{i})$ for $M$ classical coordinates $x_{i}$,
one finds that there is in general no exact stochastic process in
$(x_{i},p_{i})$ that corresponds to the quantum dynamics. One way
to understand this, is that this would be equivalent to a local hidden
variable theory, which is in general ruled out by Bell's theorem.
Despite this, these techniques are useful approximations with many
applications\cite{Drummond1993a,Norrie2005,Norrie2006,Ruostekoski}.
An alternative method is to use a nonclassical phase-space, that allows
for quantum entanglement and superpositions\cite{pp,Carusotto}.

As an example, the Positive-P representation\cite{pp} uses a non-classical
phase-space. It is able to represent all quantum density matrices,
and has a positive definite propagator. The method expands the density
matrix over coherent states $|\vec{\alpha}\rangle$ with $4M$ real
or $2M$ complex coordinates:

\begin{description}
\item 
\end{description}
\begin{equation}
\,\hat{\rho}=\int P(\vec{\alpha},\vec{\beta})\ \frac{|\vec{\beta}\rangle\!\langle\vec{\alpha}|}{\langle\vec{\alpha}|\vec{\beta}\rangle}\ d^{2M}\vec{\alpha}\, d^{2M}\vec{\beta}\end{equation}

The resulting distributions are positive and obey a diffusion equation
when the system is described by a Hamiltonian that involves single--
and two--particle processes, which can therefore be numerically simulated
on a conventional digital computer\cite{Qnoise}. A number of useful
applications of these methods to correlation function calculations
have been made\cite{Molmer,UQprl,Savage}. The trade-off is that,
when there is insufficient damping, the sampling error increases with
time, although the useful time period is easily estimated\cite{dynamix1}.
This can be controlled, to some extent, with more advanced methods
like the stochastic-gauge P-representation\cite{Deuar2002,Drummond2003,dynamix2}.

\subsection{Evaporative cooling}

The breakthrough in experimental techniques which led to the observation
of BEC was the use of evaporative cooling to produce temperatures
even lower than those obtained with laser cooling. However, as this
is a non-equilibrium process - with no conventional thermal reservoir
- it leads to an intriguing fundamental question: \emph{What is the}
\emph{true} \emph{state of an atom BEC?}

We have tried to answer this question with a first-principles quantum
simulation of the many-body quantum dynamics leading to BEC formation\cite{Corney}.
The process is simple in principle. As shown in Fig.~\ref{fig:EvapCooldiagram},
colliding warm atoms can produce a `hot' and `cold' pair: the more
energetic atom escapes, while the slow-moving partner joins the condensate.
This process is accelerated by Bose stimulation, once there are sufficiently
many particles in a low-energy single-particle state.

The system is described by the following Hamiltonian: \begin{equation}
\hat{H}=\int\,\left[\frac{\hbar^{2}}{2m}\nabla\hat{\Psi}^{\dagger}\nabla\hat{\Psi}+\frac{g}{2}\hat{\Psi}^{\dagger2}\hat{\Psi}^{2}+V(\mathbf{r},t)\hat{\Psi}^{\dagger}\hat{\Psi}+\hat{\Gamma}(\mathbf{r})\hat{\Psi}^{\dagger}+\hat{\Gamma}^{\dagger}(\mathbf{r})\hat{\Psi}\right]\, d\,^{3}\mathbf{r}.\label{H}\end{equation}
 Here $g$ represents the inter-atomic $s$-wave scattering interaction
in a point-contact approximation (with a momentum cut-off), $m$ is
the mass, $V(\mathbf{r})$ is the trap potential, and $\hat{\Gamma}(\mathbf{r})$
is a reservoir of untrapped modes which describe the spatially dependent
out-coupling from the trap. After transforming to the positive-P representation,
rescaling to a dimensionless form, sampling the distribution and obtaining
the stochastic equations for the samples, one obtains the following
equations for the complex fields $\alpha(\mathbf{z},\tau),\beta(\mathbf{z},\tau)$:
\begin{eqnarray}
\frac{d\alpha}{d\tau} & = & -i\left[\alpha\beta^{*}+v(\mathbf{z},\tau)-i\gamma(\mathbf{z})-\nabla^{2}+\sqrt{i}\zeta_{1}(\mathbf{z},\tau)\right]\alpha\nonumber \\
\frac{d\beta}{d\tau} & = & -i\left[\alpha^{*}\beta+v(\mathbf{z},\tau)-i\gamma(\mathbf{z})-\nabla^{2}+\sqrt{i}\zeta_{2}(\mathbf{z},\tau)\right]\beta,\end{eqnarray}
 where, $\zeta_{(j)}(\mathbf{z},\tau)$ are real independent white
noise fields with the variances\begin{equation}
\langle\zeta_{(i)}(\mathbf{z},\tau)\zeta_{(j)}(\mathbf{z}',\tau')\rangle=\delta_{ij}\delta(\tau-\tau')\delta^{(3)}(\mathbf{z}-\mathbf{z}')\,\,.\end{equation}

Thes equations are an exact mapping from quantum dynamics. They have
the structure of the usual Gross-Pitaevskii mean-field equations,
together with a stochastic delta-correlated field that contains the
quantum noise caused by collisions. These noise terms give rise, for
example, to the spontaneous effects that are absent in a mean-field
description. There are two coupled phase-space fields driven by independent
noises, so that eventually the equations depart from the usual classical
GP behaviour. The equations include the scaled trap potential $v(\mathbf{z},\tau)$,
and loss rate $\gamma(\mathbf{z})$.

\begin{figure}
{\small \includegraphics[width=7cm,keepaspectratio]{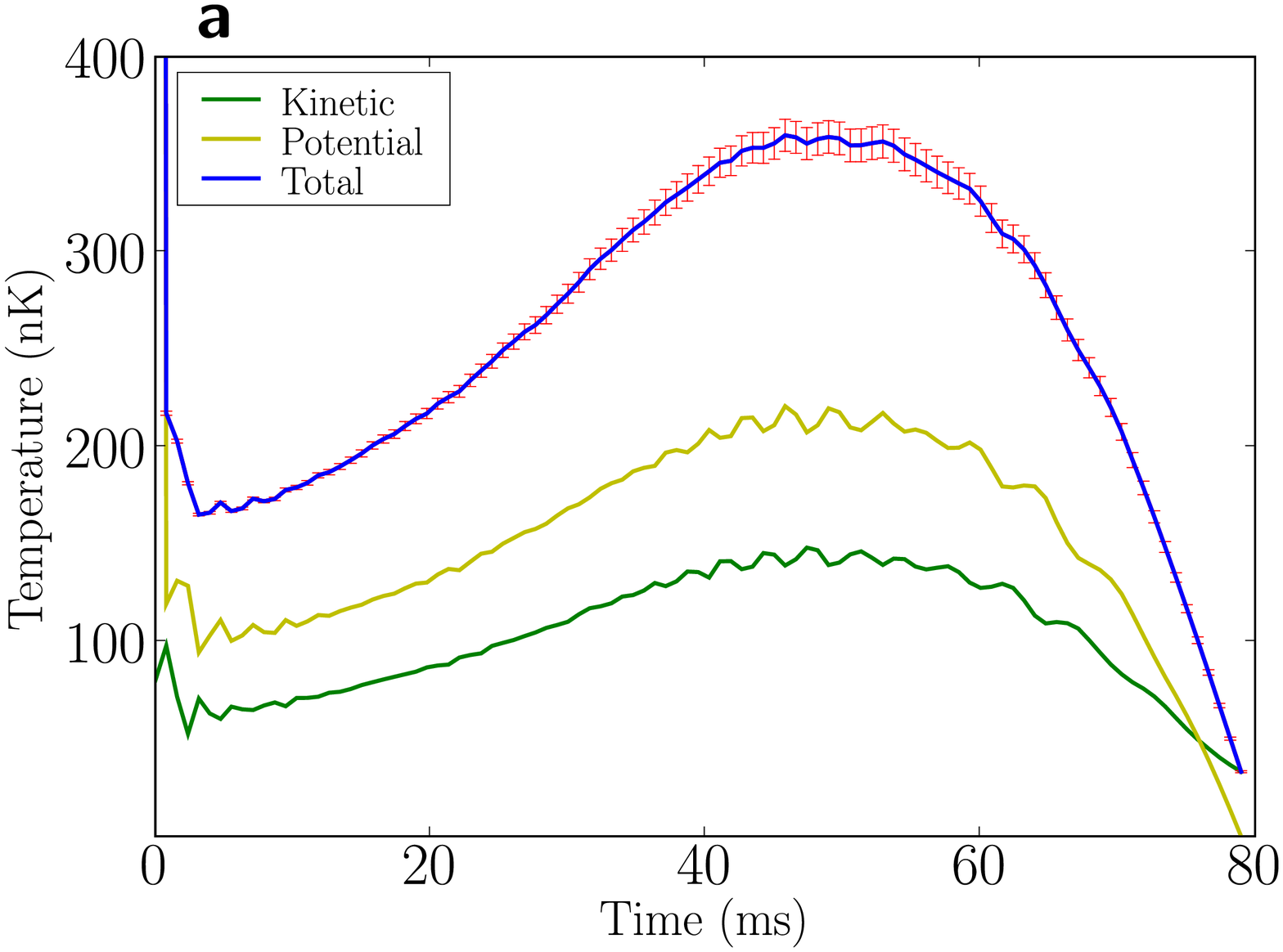}\includegraphics[width=7cm,keepaspectratio]{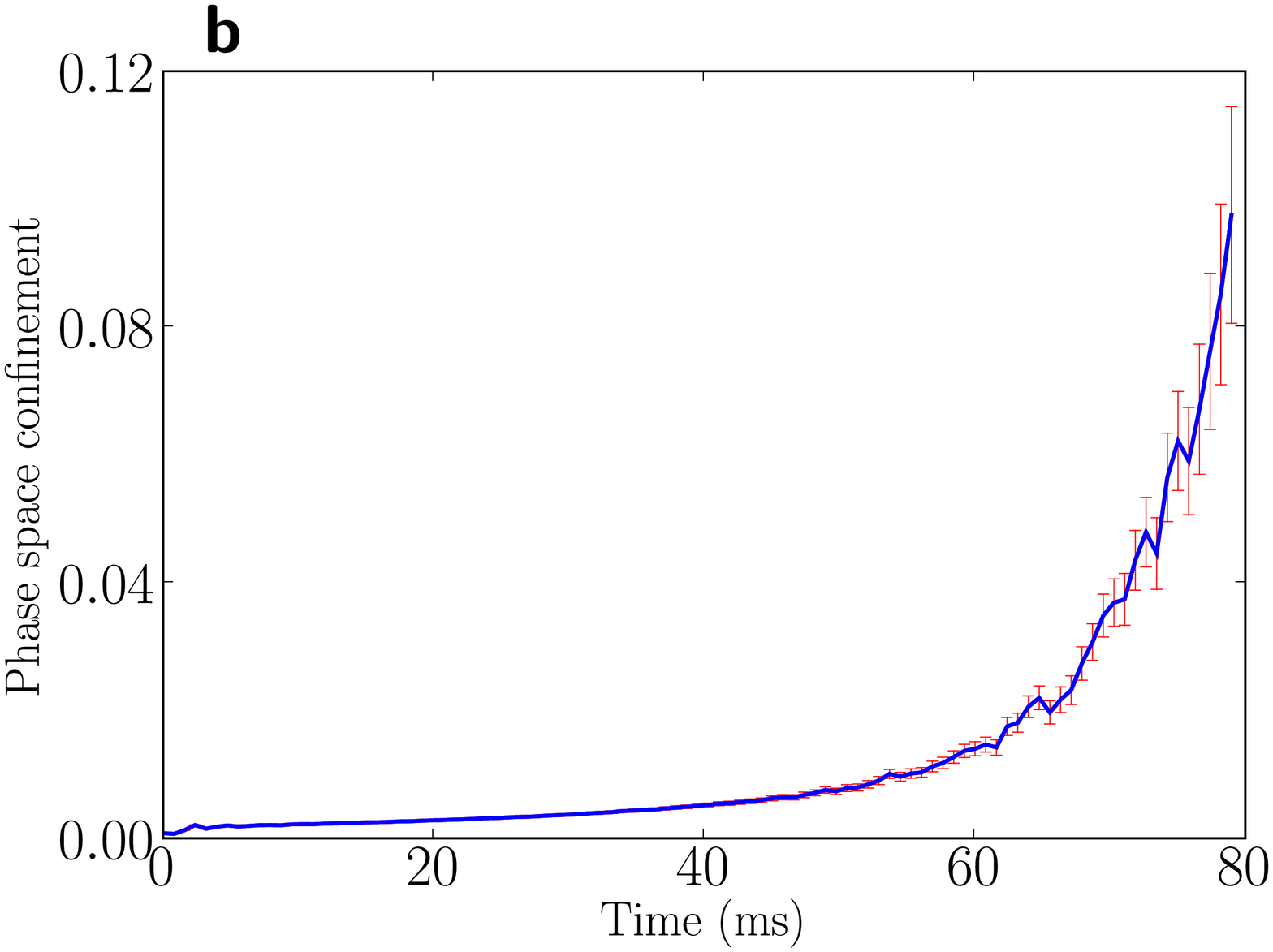}}{\small \par}

\caption{\label{fig:EvapResults} {\small Evaporative cooling simulation results:
(a) effective center-of-mass temperature and (b) second-order phase-space
filling factor ($F_{\mathbf{k}}^{(2)}$) measure of condensation.}}
\end{figure}

The results of the evaporative cooling simulations are illustrated
as Fig.~\ref{fig:EvapResults}. Firstly, Fig.~\ref{fig:EvapResults}
(a) shows the time-dependence of the center-of-mass temperature, showing
the presence of a stimulated increase in center-of-mass temperature,
during the condensate formation process. Secondly, Fig.~\ref{fig:EvapResults}
(b) illustrates the corresponding change in the second-order phase-space
filling factor $F_{\mathbf{k}}^{(2)}$ (defined in Eq.~(\ref{eq:F2})
above) during the condensate formation process, showing that this
higher-order measure of coherence can indeed provide a signature of
condensate formation.

\subsection{BEC collisions with 150,000 atoms}

\begin{figure}
\begin{centering}\includegraphics[width=10cm]{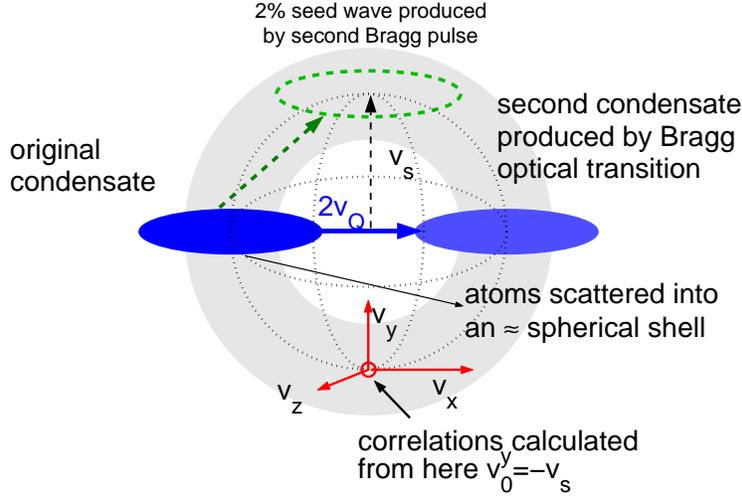} \par\end{centering}

\caption{\label{fig:schema}Schematic of the BEC collision}
\end{figure}

The collision of pure $^{23}$Na BECs, was reported in an experiment
at MIT\cite{Vogels}, and a similar setup is now being investigated
in detail with metastable He$^{*}$. This provides an analog to nonlinear
optics, in which a type of four-wave mixing occurs, generating correlated
pairs of atoms. First-principles quantum dynamical simulations can
provide a way to benchmark and test competing approximate theories,
while providing a route forward to future experimental tests of correlations.
In these simulations\cite{pdpdd2007,4wm}, schematically represented
in Fig.~\ref{fig:schema}, a $1.5\times10^{6}$ atom condensate is
prepared in a cigar-shaped magnetic trap with frequencies of $20,80,80\, Hz$
in the $X,Y,Z$ directions respectively. A brief Bragg laser pulse
coherently imparts an $X$ velocity of $2{\rm v_{Q}}\approx20$ mm/s
to half of the atoms, which is much greater than the maximum sound
velocity of $3.1$ mm/s. Simulations of both spontaneous\cite{pdpdd2007}
and stimulated\cite{4wm} scattering were carried out. For stimulated
scattering, another much weaker pulse generates a small $2\%$ \char`\"{}seed\char`\"{}
wavepacket with a $Y$ velocity of ${\rm v}_{s}=9.37$ mm/s relative
to the center of mass (this value is chosen to give strong bosonic
stimulation into the {}``fourth'' condensate).

At this point the trap is turned off so that the wavepackets collide
freely. In a center-of-mass frame, atoms are scattered preferentially
into a spherical shell in momentum space with anticorrelated directions
and mean velocities ${\rm v_{s}}\approx{\rm v_{Q}}$. Seed pulses
induce a four-wave mixing process which generates a stimulated coherent
wavepacket at $Y$ velocity -${\rm v_{s}}$, as well as growing the
strength of both of the wavepackets at $\pm{\rm v_{s}}$ by Bose enhanced
scattering.

\begin{figure}
\begin{centering}\includegraphics[width=12cm]{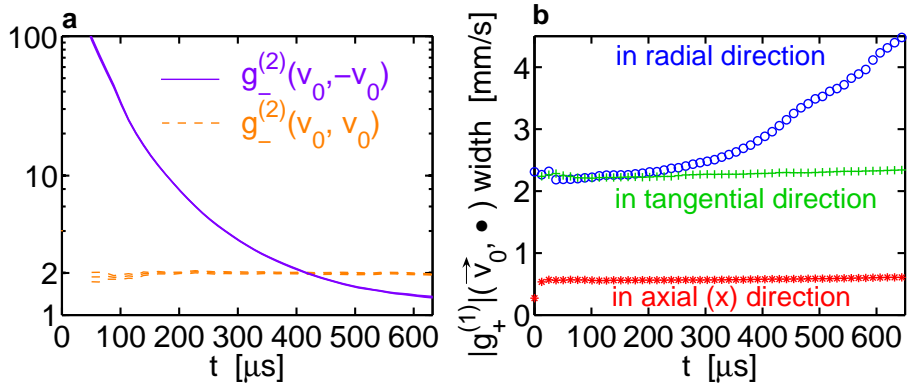} \par\end{centering}

\caption{\label{g2t}\textit{Correlations between scattered atoms: time evolution
(no seed pulse)}. Plate \textbf{a} shows the extremely strong number
correlations $g^{(2)}({\rm v}_{0},-{\rm v}_{0})$ between atoms with
opposite velocity (solid line) in the scattered shell at $|{{\rm v}_{0}}|={\rm v_{s}}$
(away from the coherent wavepackets), and thermal correlations $g^{(2)}({\rm v}_{0},{\rm v}_{0})=2$
between scattered atoms at the same velocity (dashed). Triple lines
indicate uncertainty. Plate \textbf{b} shows the coherence width in
velocity space for scattered atoms at similar velocities centered
around ${{\rm v}_{0}}$. Plotted is the Full-width at half-maximum
(FWHM) of $|g^{(1)}({\rm v}_{0},{\rm v}_{0}+{\rm v})|$. }
\end{figure}

\begin{figure}
\begin{centering}\includegraphics[width=12cm]{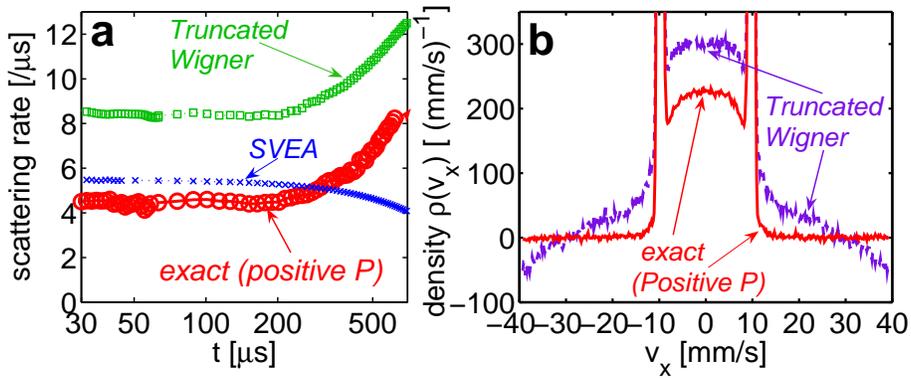} \par\end{centering}

\caption{\label{g2twig}\textit{Scattering rates: time evolution (no seed
pulse)}. Plate \textbf{a} shows total scattering rates in different
approximations compared to the exact positive-P simulations. Plate
\textbf{b} shows the distribution in velocity space for scattered
atoms, demonstrating that the false scattered halo of the truncated
Wigner method is caused by an unphysical depletion of the vacuum with
negative populations.}
\end{figure}

The dynamics of the distribution of atom velocities and correlations
between the scattered atoms have been calculated and are shown in
Figures ~\ref{g2t}.~ and~\ref{g2twig}. Such correlations have
recently become experimentally measurable\cite{Greiner2005,Folling2005,Schellekens2005}.
Correlation behaviours qualitatively similar to these have been seen
experimentally. The simulation is carried out using the positive-P
representation in the $X$-center-of-mass frame from the moment the
lasers and trap are turned off \mbox{($t=0$).} The initial wavefunction
is modeled as the coherent-state mean-field Gross-Pitaevskii (GP)
solution of the trapped $t<0$ condensate, but modulated with a factor
$\left[\sqrt{0.49}e^{im{\rm v_{Q}}{\rm x}/\hbar}+\sqrt{0.49}e^{-im{\rm v_{Q}}{\rm x}/\hbar}+\sqrt{0.02}e^{-im{\rm v_{s}}{\rm y}/\hbar}\right]$
for the case of a 2\% seed pulse, or $\sqrt{2}\,\cos(m{\rm v_{Q}x}/\hbar)$
when no seed pulse is present. The field Hamiltonian is discretized
with lattice sizes of up to $432\times105\times50$, generating a
Hubbard-type Hamiltonian like Eq (\ref{H}), but with external potential
$V$ and reservoirs $\hat{\Gamma}$ omitted.

As mentioned above, limitations arise because the size of the sampling
uncertainty grows with time, and eventually, soon after the end of
the time scale in Fig.~\ref{g2t}, reaches a size where it is no
longer practical to obtain useful precision. This time can be estimated
using the formulae found in \cite{dynamix1}, although it may be extended
with more sophisticated techniques.

A number of approximate calculations exist for this experiment\cite{svea,Bach,Zin,Norrie2005,Norrie2006},
making it useful to have a definitive first-principles approach. Comparisons
were made (Fig.~\ref{g2twig}) with previous approximate simulations
that used a scattering cross-section approximation (the SVEA method\cite{svea}),
and those\cite{Norrie2005,Norrie2006} using a classical phase-space
technique - the truncated Wigner method\cite{Drummond1993a}. This
approximate method was less accurate (as is to be expected from its
limitations when the number of particles per mode is small\cite{Sinatra})
at large momentum cutoff, due to a diverging truncation error caused
by the ultra-violet divergence of the symmetrically-ordered vacuum
fluctuations, causing an unphysical depletion of the vacuum modes
at high momentum.

The model treated up to $M=2.268\times10^{6}$ interacting momentum
modes. Since each of the $N=1.5\times10^{5}$ atoms can be in any
one of the modes, the Hilbert space contains about $N_{s}\approx M^{N}\approx10^{1,000,000}$
orthogonal quantum states. This demonstrates the ability of nonclassical
phase-space methods to simulate quantum mechanical correlations even
in the face of exceptionally large many-body Hilbert spaces.

\section{SUMMARY}

In summary, we have pointed out how pioneering twentieth century developments
of coherence and quantum optics are now growing to include ultra-cold
atomic coherence and correlations. The field has been greatly enriched
by the outstanding efforts of experimentalists to produce atomic gases
at ultra-low temperatures, and more recently, to measure their correlations.
As the field is a rapidly growing one, we cannot attempt to cover
all the new developments, but rather have provided a selection of
topics.

A simple idea treated here is to consider the standard quantum limits
to center-of mass measurements. We found that for the same density
distribution, the ground-state center-of-mass measurement variance
is $N$ times lower for a one-dimensional Fermi gas as for a Bose
gas of $N$ atoms.

Next, we briefly reviewed how coherence theory can be used to identify
Bose condensation. While the traditional Penrose-Onsager measure of
first-order coherence is very useful for a restricted class of condensates,
it fails to identify condensation within mixtures. We have shown how
higher-order coherence theory can be used to define condensation in
more general cases.

In order to illustrate how to calculate atomic correlations, we considered
the issues raised by the relatively strong inter-atomic interactions,
compared to those in quantum optics. In order to include quantum coherence
fully, we turn have turned to nonclassical phase-space methods using
the positive-P representation. This can handle three-dimensional simulations
with up to $10^{5}$ particles, and $10^{6}$ modes, giving conditions
close to those found in experiments.

Particular cases presented here include evaporative cooling and condensate
formation, which we have found is accompanied by heating of the COM
motion. We have also calculated measures of second order quantum coherence
in BEC collisions, and shown how the exact results can differ from
various approximations.

\section*{Acknowledgments}

P.~Deuar acknowledges financial support by the European Community
under the contract MEIF-CT-2006-041390. P. D. D., J.C. and T.V. acknowledge
support from the Australian Research Council Centre of Excellence
program.

\end{document}